\documentclass{PoS}

 \newcommand{\pslash}{p\kern-1ex /}
\newcommand{\lslash}{l\kern-1ex /} \newcommand{\sslash}{s\kern-1ex /}
\newcommand{\Dslash}{{\cal D}\kern-1.5ex /}

\newcommand{\beqa}{\begin{eqnarray}}
\newcommand{\eeqa}{\end{eqnarray}} \newcommand{\be}{\begin{equation}}
\newcommand{\ee}{\end{equation}} \newcommand{\bea}{ \begin{eqnarray}}
\newcommand{\eea}{\end{eqnarray}}
\newcommand{\ba}{\begin{array}}
\newcommand{\ea}{\end{array}}

\newcommand{\pref}[1]{(\ref{#1})}

\newcommand{\Tr}[1]{\langle {#1} \rangle}
\newcommand{\csw}{\overline{c}_{{\rm SW}}}
\newcommand{\ca}{\overline{c}_{{\rm A}}}
\newcommand{\cv}{\overline{c}_{{\rm V}}}

\title{The vector and axial currents in Wilson chiral perturbation theory}

\ShortTitle{The vector and axial currents in Wilson chiral perturbation theory}

\author{Sinya Aoki\\
Graduate School of Pure and Applied Sciences,
  University of Tsukuba, Tsukuba, Ibaraki 305-8571, Japan\\
E-mail: \email{saoki@het.ph.tsukuba.ac.jp}}

\author{Oliver B\"ar\\
Institute of Physics, Humboldt Universit{\"at} zu Berlin,
  Newtonstra{\ss}e 15, 12489 Berlin, Germany\\
Email: \email{obaer@physik.hu-berlin.de}}

\author{\speaker{Stephen R. Sharpe}\\
        Physics Department University of Washington Seattle, WA 98195-1560\\
        E-mail: \email{sharpe@phys.washington.edu}}

\abstract{
We reconsider the construction and matching of the vector and axial currents
in Wilson Chiral Perturbation Theory (WChPT), the low-energy effective
theory for lattice QCD with Wilson fermions.  In particular, we discuss 
in detail the impact of the 
finite renormalization of the currents 
on their matching from the lattice theory to WChPT. 
We explicitly show
that imposing chiral Ward-Takahashi  
identities on the currents leads, in general, 
to additional terms of O($a$) in the axial current.
We illustrate the impact on physical quantities 
by computing the pion decay constant to one-loop order in the two flavor
theory. Our result differs from previously published ones. 
}

\FullConference{The XXVII International Symposium on Lattice Field Theory\\
                 July 26-31, 2009\\
                 Peking University, Beijing, China}

\begin{document}

\section{Introduction}
Lattice QCD breaks several global
symmetries of the (formal) continuum theory.
It follows that
the corresponding lattice currents are not exactly conserved,
and require finite renormalization to match to continuum currents.
A standard method for determining the renormalization factors is to
impose Ward-Takahashi identities (WTI)
which would hold were the symmetry exact.
The resulting renormalization factors depend on the WTI chosen,
although this dependence vanishes with some power of the lattice spacing.

An important example is the breaking of 
chiral symmetry when using Wilson fermions (possibly improved).
The current is then the non-singlet axial
current, and one determines $Z_{\rm A}$ by imposing a WTI following from
$SU(2)_L\times SU(2)_R$ chiral symmetry 
(assuming two flavors)~\cite{Luscher:1996sc}. 

In this note we address the question of how $Z_{\rm A}$ depends upon
the choice of WTI, and how the dependence enters into results for
matrix elements of the renormalized axial current, e.g. $f_\pi$.
As noted already, the variation in $Z_{\rm A}$ is a discretization error,
and thus the question is naturally addressed using Wilson chiral
perturbation theory (WChPT), i.e. ChPT with discretization errors
incorporated~\cite{Sharpe:1998xm}.
It turns out, however, that the results on this issue in the 
literature~\cite{Rupak:2002sm,Sharpe:2004ny} are incorrect.
Here we present a brief summary of our recent 
reanalysis~\cite{Aoki:2009ri}, 
explaining what was missed in the earlier work, 
and commenting on the significance of our new results.
Our analysis controls only discretization errors linear in $a$,
and thus is useful only if either the action or the currents (or both)
are unimproved.
The methodology can, however, be extended to higher order in $a$,
and the overall conclusions are applicable more generally.

We also discuss, in tandem, the corresponding issue for ultra-local 
lattice vector currents. Here there {\em is} an exactly conserved
lattice current, but it is not ultra-local, and
often it is computationally simpler to use the ultra-local version.
Since the latter current is not conserved, it receives a finite
renormalization, one that can again be determined by imposing an
appropriate WTI.

\section{Currents and Renormalization Conditions}

We consider lattice QCD with $N_{f}=2$ Wilson fermions 
with lattice spacing $a$. 
The ultra-local flavor non-singlet vector and axial currents
are related to renormalized currents as follows:
\bea 
V_{\mu,{\rm ren}}^{b} = Z_{\rm V} 
\ V_{\mu,{\rm Loc }}^{b}
\qquad
\label{eq:RenLocV}
&& V_{\mu,{\rm Loc}}^{b}(x)\, = \, \overline{\psi}(x)
\gamma_{\mu}T^{b}\psi(x)
\label{eq:RenV}
\\ 
A_{\mu,{\rm ren}}^{b} = Z_{\rm A} 
\ A_{\mu,{\rm Loc}}^{b}  \label{eq:RenLocA} 
\qquad 
&&
A_{\mu,{\rm Loc}}^{b}(x)\,=\, \overline{\psi}(x)
\gamma_{\mu}\gamma_{5}T^{b}\psi(x)  
\label{eq:RenA} 
\eea
The $Z$-factors depend on the action as well as 
on the WTI used to fix them.
For simplicity, we work here and below in the massless
limit, attained by sending $\kappa\to\kappa_c$.
Since we work only at linear order in $a$, we do not encounter
the possible Aoki-phase or first-order transition at small
masses induced by discretization errors of $O(a^2)$~\cite{Sharpe:1998xm}.
This simplification also 
means that we drop all terms proportional
to $am$.\footnote{%
In practice, attaining zero quark mass requires an extrapolation,
or the use of Schr\"odinger-functional boundary conditions as
an infrared regulator.}

The specific WTI that we use to determine the $Z$-factors are as follows.
For the vector current, we impose that the pion 
have the correct ``charge'':
\bea
\label{eq:ZVLatt2} 
\langle \pi^{b}(\vec{p}) 
| V^{c}_{0,{\rm ren}} | \pi^{d}(\vec{p}) \rangle& = & \epsilon^{bcd} 2 E 
\eea
One must have that $\vec p\ne0$ in order that $E\ne 0$ (since we
are in the massless limit).
The renormalization factor will, in general, depend on $\vec p$.

For the axial current we impose ``$\delta_A A\sim V$''
between pion states~\cite{Luscher:1996sc}
\bea
\label{eq:AVWI3} 
\int {\rm d}\vec{x}\, \epsilon^{abc}\epsilon^{cde} \langle
\pi^{d}(\vec{p})| \left[A_{0,{\rm ren}}^{a}(y_{0}+t,\vec{x}) - 
  A_{0,{\rm ren}}^{a}(y_{0}-t,\vec{x})\right]\, A_{0,{\rm ren}}^{b}(y) |
\pi^{e}(\vec{q})\rangle \nonumber\\
= 2i \epsilon^{cde} \langle
\pi^{d}(\vec{p})| V_{0,{\rm ren}}^{c}(y) |\pi^{e}(\vec{q}) \rangle\,.
\eea 
Here we need either $\vec p$ or $\vec q$ (or both) to be non-vanishing 
for the right-hand-side to be non-zero, and we choose to
take both non-vanishing to avoid infrared divergences. 
Once we have a renormalized vector current (which enters on the 
right-hand-side) we can use this WTI to determine $|Z_{\rm A}|$.
Here the result can depend not only on the pion momenta
but also on the Euclidean time separating
the two axial currents (which traces back to the size of the
region over which the axial rotation is applied).

Other renormalization conditions are sometimes used in practice,
but we have chosen these two examples as they can be analyzed using ChPT.
We also note that, in practice, one cannot consider many
choices for the external pion momenta and time separations,
due to the associated computational cost and the degradation of the signal
as $|\vec p|$ increases. In contrast, ChPT allows one to study
all momenta in the chiral regime.

\section{Mapping Currents into the Effective Chiral Theory}

The required calculation is now clear: map the lattice currents
into the effective chiral theory and then evaluate the matrix
elements appearing in the WTI (\ref{eq:ZVLatt2}) and (\ref{eq:AVWI3}).
The mapping is done, following~\cite{Sharpe:1998xm},
in two steps, first from the lattice into the Symanzik
effective continuum theory, and from there into the chiral effective
theory.
Since $Z_{\rm V}$ and $Z_{\rm A}$ 
are overall constants to be determined at the end,
the mapping needs to be done for the bare lattice currents.

In the first step, using the
symmetries of the lattice theory, one finds~\cite{Luscher:1996sc}:
\bea 
V_{\mu,{\rm Loc}}^{b} &\simeq&
V_{\mu,{\rm Sym, Loc}}^{b} = 
\frac1{Z_{\rm V}^0}
\left(
V_{\mu,{\rm ct}}^{b} + a \cv \partial_{\nu} T^{b}_{\mu\nu,{\rm ct}}
\right) + {\rm O}(a^{2})
\label{eq:SymVC}
\\ 
A_{\mu,{\rm Loc}}^{b} &\simeq & 
A_{\mu,{\rm Sym, Loc}}^{b} = 
\frac1{Z_{\rm A}^0}
\left(
A_{\mu,{\rm ct}}^{b} + a \ca \partial_{\mu} P_{\rm  ct}^{b}
\right) + {\rm O}(a^{2})
\label{eq:SymAC} 
\eea
where the continuum bilinears ${\cal O}_{\rm ct}$ take their usual forms.
This result displays all power-law dependence on $a$ explicitly
(here at linear order);
a logarithmic dependence still enters through the implicit dependence of
$\overline{c}_{\rm V,A}$ and $Z_{\rm V,A}^0$ on $g(a)$.
$Z$-factors are needed in these mappings because
the currents $V_{\mu,\rm ct}^a$ and $A_{\mu,\rm ct}^a$ are conserved
while the lattice currents are not. 
The superscript indicates that the $Z_{\rm V,A}^0$, while containing
perturbative contributions to all orders, do not include $O(a)$ terms.
In practice, we know $Z_{\rm V,A}^0$ only approximately, but this does
not matter since, as will be seen below, 
they cancel once we normalize the currents non-perturbatively.

The mapping of the currents in (\ref{eq:SymVC}-\ref{eq:SymAC})
into Wilson ChPT is more subtle. One cannot simply take the
WChPT action (including $O(a)$ terms)
and determine the currents using the Noether procedure,
as done in Ref.~\cite{Rupak:2002sm}, because the symmetries are broken
explicitly. Instead, we have used the following two methods.
\begin{itemize}
\item Following continuum ChPT, construct the
generating functional, including lattice artefacts
using standard spurion methods, and obtain the currents by taking
derivatives with respect to sources. One must treat the
$O(1)$ and $O(a)$ parts of the currents (\ref{eq:SymVC}-\ref{eq:SymAC})
separately.
\item  Write down the most
general currents compatible with the symmetries, and impose
appropriate WTI 
 by hand.  
\end{itemize}
The first approach was used in Ref.~\cite{Sharpe:2004ny},
and more fully justified in \cite{Sharpe:2006pu}. We have
checked the analysis using a more general method. 
The second approach is more direct, and we used it to
check that no subtleties had been overlooked in the first method.
We find that both approaches agree.

The resulting mappings are (in agreement with the results of
Ref.~\cite{Sharpe:2004ny}, 
and dropping here and henceforth a ubiquitous ``$+ O(a^2)$''): 
\be 
V_{\mu,{\rm ct}}^{b} \!+\! a \cv \partial_{\nu} T^{b}_{\mu\nu,{\rm ct}} 
\simeq
V_{\mu, {\rm eff}}^{b}  = V_{\mu,{\rm LO}}^{b}\left(1 +
\frac{4}{f^{2}}\hat{a}{W}_{45} \csw 
\Tr{\Sigma \!+\!  \Sigma^{\dagger}}\right)\,,
\label{eq:VeffNoether} 
\ee
\be
A_{\mu,{\rm ct}}^{b}\! +\! a \ca \partial_{\mu} P_{\rm  ct}^{b} 
\simeq
A_{\mu,{\rm eff}}^{b} = A_{\mu,{\rm
    LO}}^{b}\left(1\! +\! \frac{4}{f^{2}}\hat{a}{W}_{45}\csw 
\Tr{\Sigma \!+\!   \Sigma^{\dagger}} \right) 
+ 4\hat{a}{W}_{\rm A}\ca\partial_{\mu}
\Tr{T^{b}(\Sigma \!-\! \Sigma^{\dagger})}, 
\label{eq:Adirect2}
\ee
where ${V}_{\mu,{\rm LO}}^a$ and ${A}_{\mu,{\rm LO}}^a$ are the standard, 
leading-order ChPT currents,
$\Sigma$ contains the pion fields in the standard way, 
$\hat{a}=2W_0 a$, and the $W_X$ are unknown low energy coefficients
(LECs) associated with the lattice artifacts.
The coupling $\csw$ is the coefficient of the ``clover-term''
in the Symanzik action, and, like $\overline{c}_{\rm V,A}$, is only
known approximately. This is now seen to be unimportant, however,
since it multiplies an unknown LEC.
Note that for the vector current the $\cv$ term in the Symanzik-theory
current does not lead to an $O(a)$ contribution to the current in WChPT,
while the $\ca$ term in the axial current does. 

Putting in the overall factors, 
the renormalized ultra-local vector and axial currents map into WChPT as
\begin{equation}
V_{\mu,\rm ren}^b \simeq
\frac{Z_{\rm V}}{Z_{\rm V}^0} 
V_{\mu,\rm eff}^b\,,\qquad\qquad
A_{\mu,\rm ren}^b \,\simeq\,
\frac{Z_{\rm A}}{Z_{\rm A}^0} 
A_{\mu,\rm eff}^b\,.\label{eq:VAlocWChPT}
\end{equation}
In Ref.~\cite{Sharpe:2004ny}, the factors of $Z_{\rm V}/Z_{\rm V}^0$
and $Z_{\rm A}/Z_{\rm A}^0$ were set to unity, based on an erroneous
argument. In fact, these factors can differ from unity at O($a$), as
will be seen shortly.

\section{Determining the Renormalization Factors}

We now use the mapped currents of
(\ref{eq:VAlocWChPT})
to evaluate the WTI in WChPT. We work
in the power-counting in which $a\sim p^2$, so that a tree-level
calculation suffices to obtain the terms linear in $a$.
Evaluating the WTI (\ref{eq:ZVLatt2}) we find the simple result
\bea
\label{eq:ZV} 
Z_{\rm V} & = & Z_{\rm V}^0
\qquad V_{\mu,\rm ren}^b = V_{\mu,\rm eff}^b\,. 
\eea
Hence, no additional O($a$) terms are 
introduced by the current renormalization. We discuss why
this is in Ref.~\cite{Aoki:2009ri}; see also Ref.~\cite{Sharpe:2004ny}. 

For the axial current, the WTI (\ref{eq:AVWI3}) gives, after some
calculation,
\be
\frac{Z_{\rm A}}{Z_{\rm A}^0} 
=
1 - \frac{4\hat{a}}{f^2}(W_{45}\csw+W_{\rm A}\ca) 
z_{\rm A}(t)\,,\label{ZAresult}
\ \ z_{\rm A}(t) \,=\,
1 - \cosh[t(|\vec p|-|\vec q|)] \exp[-|t| |\vec p - \vec q|]\,.
\label{eq:zAt}
\ee 
In contrast to the vector current we do find a non-vanishing correction
of O($a$). This correction depends 
on the separation $t$ between the axial currents and upon the 
external states, as expected. Thus, the renormalized ultra-local axial
current maps into WChPT as
\begin{eqnarray}
A_{\mu,\rm ren}^b &\simeq& 
\left[1 - \frac{4\hat{a}}{f^2}(W_{45}\csw+W_{\rm A}\ca) z_{\rm A}(t)\right]
A_{\mu,\rm eff}^b\,.
\label{eq:AlocWChPT_final}
\end{eqnarray}
This is our major new result.

We conclude that the $a$ dependence of $A_{\mu,\rm eff}^b$, 
derived using symmetries, is supplemented by
an additional discretization error resulting from the
application of the normalization condition at non-zero $a$.
There are three distinct cases 
(recalling that $\vec p,\vec q\ne 0$):
\begin{enumerate}
\item
$\vec p=\vec q$. This is the simplest case to implement
practically, and leads to $z_{\rm A}(t)=0$. Thus it turns out that,
in this case,
there are no additional $O(a)$ terms introduced by the current
normalization.
\item
$\vec p$ parallel to $\vec q$. Then, for $|t|\gg 1/|\vec p-\vec q|$,
the product of cosh and exponential becomes $1/2$, and so
$z_{\rm A}\to 1/2$.
\item
All  other non-vanishing $\vec p$ and $\vec q$.
Here, for $|t|\gg 1/|\vec p-\vec q|$, the exponential
overwhelms the cosh and $z_{\rm A}\to 1$.
\end{enumerate}
We stress that in both the second and third cases $z_{\rm A}$
depends on $t$ for non-asymptotic values of $t$.
In any case, our main point is that the implementation
of the renormalization condition leads, in general,
to a non-trivial O($a$) correction to the current.
This is what was missed in Ref.~\cite{Sharpe:2004ny}.  

Another feature of the result (\ref{eq:AlocWChPT_final})
is that it depends on the same unknown coefficients as
appear in the unnormalized current (\ref{eq:Adirect2}),
namely the products $W_{45}\csw$ and $W_{\rm A} \ca$.
Thus if one were to do a fit to multiple physical
quantities, incorporating the constraints implied by WChPT
so as to improve the extrapolation to the continuum limit,
the inclusion of the correctly normalized axial current
would not increase the number of unknown parameters.

\section{Applications}

Expanding the renormalized axial current
we obtain the tree-level result for the pion decay constant:
\bea
\label{eq:fpiLO}
f_{\pi,{\rm tree}} 
& = & 
f\left(1 + \frac{4}{f^{2}} \hat{a} 
\left(W_{45}\csw + W_{\rm A}\ca\right)
\left[2 - z_{\rm A}(t)\right]\right) \,.
\eea 
This gives the form of the discretization errors expected
in a lattice calculation of $f_\pi$.
We stress once again that the result depends
on the choice of renormalization condition [through $z_{\rm A}(t)$].
While expected, it is still striking to see such
a dependence explicitly. 

Various comments are in order. First, at this order in ChPT, the
continuum result is simply $f$, which is correctly reproduced.
Second, a consistency condition on the calculation is that the
LECs can appear in physical quantities only in 
certain ``physical''
combinations~\cite{Sharpe:2004ny}, and
$W_{45}\csw+W_{\rm A}\ca$ is indeed such a combination.
Third, the choice of underlying fermion action enters through the
values of $\csw$ and $\ca$, with the choice of the ultra-local
current also affecting $\ca$. If only the action is improved then
$\csw=0$, and one still finds, as expected, an O($a$) term because
$\ca\ne0$. If both action and current are improved, then one finds
the expected absence of the O($a$) term---independent of the choice
of the details of the renormalization condition.
Finally, we note that in comparison with previous results in
the literature, Ref.~\cite{Rupak:2002sm} misses both $W_{\rm A}$ 
and $z_{\rm A}$ terms, while
Ref.~\cite{Sharpe:2004ny} effectively assumes that $z_{\rm A}=0$.

We have also done a one-loop calculation and find
\bea\label{eq:fpiNLO} 
f_{\pi,{\rm 1-loop}} & = & f\left(1 +
\frac{\hat{a}}{f^2} \tilde W_{\rm A1} - \frac{1}{16\pi^{2}f^{2}}\left[1 +
  \frac{\hat{a}}{f^2}\tilde{W}_{\rm A2}\right]
M^{2}_{\pi}\ln\frac{M^{2}_{\pi}}{\mu^{2}} + \frac{8}{f^{2}}
M^{2}_{\pi} \left[L_{45} + \frac{\hat{a}}{f^2}\tilde{W}_{\rm
    A3}\right]\right)
\eea
with $\tilde{W}_{\rm A3}$ being a new unknown LEC while     
    \bea    
\tilde{W}_{\rm A1} & = & 4(W_{45}\csw + W_A\ca)[2-z_A(t)]\qquad
\tilde{W}_{\rm A2} \,=\, 4(W_{45}\csw +W_{A}\ca)[1-z_A(t)]
\eea
Note that these two coefficients depend only on the physical
combination $W_{45}\csw + W_A\ca$ of LEC, as expected.

Our one-loop result correctly reproduces the continuum result of
Gasser and Leutwyler \cite{Gasser:1983yg} in the limit $a\rightarrow
0$.  At non-zero lattice spacing, however, there appear additional
terms of O($a$), O($aM_{\pi}^{2}$) and O($aM_{\pi}^{2}\ln
M_{\pi}^{2}$). Quite generally, the coefficient of the chiral
logarithm receives a correction in form of the factor $[1 +
\hat{a}\tilde{W}_{\rm A2}/f^2]$, so it not only depends on $f$ and the
number of flavors, but also on the (non-universal) lattice artifacts
encoded in the coefficient $\tilde{W}_{\rm A2}$~\cite{Aoki03}. 
An exception is the
third case discussed in the previous section with $z_A(t)=1$. Here
$\tilde{W}_{\rm A2}=0$ and the chiral logarithm is free of O($a$)
corrections.

Note that the combination $L_{45}$ of Gasser-Leutwyler coefficients
enters the one-loop result in form of the lattice spacing dependent
combination $L_{45}^{\rm eff}(a)=L_{45} + \hat{a} \tilde{W}_{\rm
A3}/f^2$. In order to obtain the physically interesting part $L_{45}$
one has to extrapolate to the continuum limit.

We close with a final remark on formula (\ref{eq:fpiNLO}). For a
consistent result to a given order one has to specify a power-counting
scheme, and the literature usually distinguishes two regimes: (i) the
GSM regime with $m\sim a\Lambda_{\rm QCD}^2$ and (ii) the LCE regime
with $m\sim a^2\Lambda_{\rm QCD}^3$. Here GSM stands for {\em
generically small quark masses} \cite{Sharpe:2004ny} and LCE for {\em
large cut-off effects} \cite{Aoki:2008gy}. Equation (\ref{eq:fpiNLO})
is only a partial next-to-leading order (NLO) 
result for the LCE regime since we ignored the
O($a^2$) corrections in the effective action \cite{Bar:2003mh} and the
effective currents. The NLO result for the GSM regime, however, is
obtained from \pref{eq:fpiNLO} by dropping the corrections
proportional to $\tilde{W}_{\rm A2}$ and $\tilde{W}_{\rm A3}$, which are of
next-to-next-to-leading order in this regime.

\section{Conclusions}

We have reconsidered the construction and matching of the vector and axial
currents in WChPT. The explicit chiral symmetry breaking
of Wilson fermions compels us to take two aspects into
account which are not present in continuum ChPT:

\begin{enumerate}
\item The local lattice currents are not conserved, and in general
they do not map onto the conserved currents in WChPT. In particular,
the WChPT currents are not obtained by the Noether procedure, because the currents in the Symanzik theory have O($a$) corrections which are not related to the effective action.
\item The proper matching of the currents has to take into account the
finite renormalization of the local lattice currents. 
The same renormalization conditions that
have been employed for the lattice currents must be imposed on the
effective currents in WChPT. 
Depending on the particular choice for the renormalization conditions
the expressions for the renormalized currents differ by terms of
O($a$).
\end{enumerate} 
A direct consequence is that WChPT predictions for matrix elements of the
currents also depend on the  renormalization
condition one has adopted. This is not a flaw but rather reflects the fact that the
lattice data differs too depending on the particular condition one has chosen.
What we find is that this dependence enters at O($a$), 
but that it does not introduce any new LECs. 

Our results emphasize the perhaps rather
obvious general point that non-perturbative renormalization conditions
generically introduce additional discretization errors.
These must be (and in practice usually are being)
accounted for when extrapolating to the continuum limit.

\section*{Acknowledgments}
This work is supported in part by the Grants-in-Aid for Scientific
Research from the Ministry of Education, Culture, Sports, Science and
Technology (Nos. 20340047,20105001,20105003), by the Deutsche
Forschungsgemeinschaft (SFB/TR 09) and by the U.S. Department of
Energy under grant no. DE-FG02-96ER40956.


\begin{thebibliography}{99}
\bibitem{Luscher:1996sc} 
  M.~L{\"u}scher, S.~Sint, R.~Sommer and P.~Weisz,
  Nucl.\ Phys.\  B {\bf 478}, 365 (1996)
  [arXiv:hep-lat/9605038].

\bibitem{Sharpe:1998xm}
  S.~R.~Sharpe and R.~L.~Singleton,
  Phys.\ Rev.\  D {\bf 58}, 074501 (1998)
  [arXiv:hep-lat/9804028].

\bibitem{Rupak:2002sm}
  G.~Rupak and N.~Shoresh,
  Phys.\ Rev.\  D {\bf 66}, 054503 (2002)
  [arXiv:hep-lat/0201019].

\bibitem{Sharpe:2004ny} 
  S.~R.~Sharpe and J.~M.~S.~Wu,
  Phys.\ Rev.\  D {\bf 71}, 074501 (2005)
  [arXiv:hep-lat/0411021].

\bibitem{Aoki:2009ri} 
  S.~Aoki, O.~B{\"a}r and S.~R.~Sharpe,
  Phys.\ Rev.\  D {\bf 80}, 014506 (2009)
  [arXiv:0905.0804 [hep-lat]].

\bibitem{Sharpe:2006pu}
  S.~R.~Sharpe,
  in ``Perspectives in Lattice QCD,'', ed. Y. Kuramashi,
  World Scientific (2008),
  arXiv:hep-lat/0607016.
  
\bibitem{Gasser:1983yg} J.~Gasser and H.~Leutwyler, \newblock
  Ann. Phys. {\bf 158} (1984) 142.
  
\bibitem{Aoki03} S.~Aoki,
  Phys.\ Rev.\ D {\bf 68}, 054508 (2003)
  [arXiv:hep-lat/0306027].

\bibitem{Aoki:2008gy}
  S.~Aoki, O.~B{\"a}r and B.~Biedermann,
  Phys.\ Rev.\  D {\bf 78}, 114501 (2008)
  [arXiv:0806.4863 [hep-lat]].

\bibitem{Bar:2003mh}
  O.~B{\"a}r, G.~Rupak and N.~Shoresh,
  Phys.\ Rev.\  D {\bf 70}, 034508 (2004)
  [arXiv:hep-lat/0306021].
  
\end{thebibliography}
\end{document}